# Parametric metasurfaces for electromagnetic wave amplification

FEDOR V. KOVALEV* AND ILYA V. SHADRIVOV

*ARC Centre of Excellence for Transformative Meta-Optical Systems (TMOS), Research School of Physics, The Australian National University, Canberra ACT 2601, Australia*
*fedor.kovalev@anu.edu.au*

**Abstract:** We study parametric amplification of electromagnetic waves using metasurfaces. We design a variable capacitor-loaded metasurface that can amplify incident electromagnetic waves. We analyze various regimes of operation of the system and find that we can achieve a significant gain (over 10 dB) in just one layer of such a structure, and this gain can be controlled by parametric modulation. We study the instability threshold for this system and show that a simple theoretical model agrees well with the results of full numerical simulations.

## 1. Introduction

Metasurfaces are composite structures whose elements, or meta-atoms, are smaller than the wavelength of electromagnetic waves of interest. If parameters of metasurfaces change over time, they are called parametric [1]. There is a range of approaches towards creating metasurfaces whose properties change over time: they can be tuned either electrically, mechanically, optically, or thermally. Nonlinear mechanisms, as well as magnetic and chemical approaches, have also been used to control the characteristics of metasurfaces. The choice of tuning technology depends on various features of metasurface design, such as the required speed of modulation and local access to individual meta-atoms [2]. Spatiotemporal metasurfaces, in which parameters are controlled both in time and space, are attracting growing interest in the research community [3–6].

Parametric metasurfaces, the properties of which depend on their modulation over time, provide access to many non-trivial physical effects. For example, they can shift the frequency of radiation [7–11], steer electromagnetic beams [12,13], control parametric waves in the scattered field [14], and exhibit nonreciprocity [15–19], meaning that electromagnetic waves are transmitted through the metasurface differently depending on the direction of propagation [20,21]. In general, achieving a higher modulation rate corresponds to unlocking more intriguing effects [2], such as compression and spectral manipulation of pulses [22], as well as parametric amplification discussed in this work. With higher modulation speeds, it is possible to achieve complete control over light both in space and time [23].

The scattering of electromagnetic waves can be controlled by the independent modulation of the electric and magnetic polarizations of metasurfaces over time [24]. Such inhomogeneous metasurfaces can achieve nearly arbitrary control over the amplitudes and phases of generated parametric waves. Given that this control can be independently implemented in each meta-atom, it becomes possible to generate waves, such as those with phase gradients, leading to controlled steering or focusing/defocusing of the waves. In addition, complex phase gradients allow metasurfaces to simultaneously function as different devices for different sidebands.

Recently, several research groups used various methods to amplify incident radiation with the help of metasurfaces, such as rotation of the anisotropy of surface susceptibility [25], utilization of active substrates [26], permittivity modulation [27,28], and integration of power amplifiers [29,30] or transistors [31,32]. Parametric amplification in left-handed transmission lines [33–35] and metamaterial waveguides [36,37] was studied at the beginning of this century, but these methods have not yet been applied to the amplification of free space electromagnetic waves. Parametric amplification techniques were originally used in electronic circuits and became the leading technology for ultralow-noise microwave measurements in quantum computing [38].

## 2. Theoretical model

In this study, we propose a new approach to electromagnetic wave amplification using metasurfaces. We choose a split-ring resonator (SRR) as the model element of our structure. By adding a modulated variable capacitor to an SRR, we convert it to a parametric amplifier. When the capacitance of a varactor undergoes periodic changes at a frequency approximately twice that of the incident wave, according to the theory of parametric amplifiers, we anticipate a unidirectional influx of energy into an SRR.

Figure 1 shows the developed parametric metasurface (a) and a lumped element circuit diagram of its meta-atom (b). The meta-atom includes the copper SRR placed on the substrate (Rogers RO4003C) and the embedded varactor diode (SMV1405), which is modulated using the pump sources via the filtering circuit ($C_F$ = 10 pF, $L_F$ = 650 uH, $R_F$ = 200 kΩ). The SRR parameters are as follows: $r$ is the SRR inner radius, $w$ is the SRR track width, $g$ is the SRR gap size and $h$ is the SRR track height. The gap size is selected to accommodate the varactor diode. The substrate thickness is 508 μm and the unit cell size is 34 mm × 34 mm.

The schematic in Figure 1(b) shows the effective circuit model for the SRR loaded by a varactor with a biasing network. The capacitor $C_f$ together with the resistor $R_f$ (a high-pass filter) protects the AC pump source from the DC voltage $U_0$ which is used to set the operating bias voltage on the capacitance-voltage characteristic of the varactor diode C(t). The inductor $L_f$ along with the resistor $R_f$ (a low-pass filter) is used to protect the DC voltage source from the AC currents at $\omega_s$ and $\omega_p = 2\omega_s$.

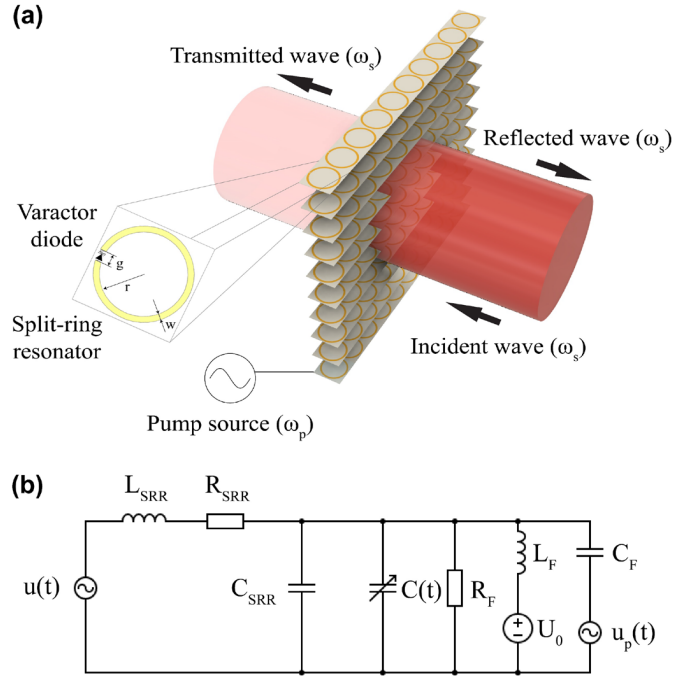

Fig. 1. (a) Schematics of the parametric metasurface made of one layer of resonators loaded with variable capacitance diodes. (b) The lumped element diagram of the meta-atom.

In this work we used the following formulas to calculate the capacitance $C_{SRR}$ and inductance $L_{SRR}$ of the SRR [39]:

$$C_{SRR} = C_{gap} + C_{surf}, \quad (1)$$

$$C_{gap} = \varepsilon_0 \left(\frac{hw}{g} + h + w + g\right), \quad (2)$$

$$C_{surf} = \frac{2\varepsilon_0(h+w)}{\pi} \log\left[\frac{4\left(r_m - \frac{w}{2}\right)}{g}\right], \quad (3)$$

$$L_{SRR} = \mu_0 r_m \left[\log\left(\frac{8r_m}{h+w}\right) - \frac{1}{2}\right], \quad (4)$$

where $r_m = r + \frac{w}{2}$ is the SRR average radius, $\varepsilon_0$ and $\mu_0$ is the free-space permittivity and permeability, respectively. The effective resistance $R_{SRR}$ in Figure 1b includes both radiation and loss resistance of the split-ring resonator.

Next, we consider in detail the SRR with an integrated varactor, which is excited via the electromotive force induced by the incident wave in the following way:

$$u(t) = U_s \cos\omega_s t, \quad (5)$$

where $\omega_s$ is the signal frequency equal to the SRR resonance frequency, $u(t) = -\frac{d\Phi}{dt}$ is the induced electromotive force in the SRR, and $\Phi$ is the magnetic flux through the SRR.

Modulation of the varactor response is performed by an independent source with a voltage given by

$$u_p(t) = U_0 + U_p \cos(\omega_p t + \varphi_p), \quad (6)$$

where $U_0$ is the constant bias voltage, $U_p$ is the pump amplitude, $\omega_p$ is the pump frequency, $\varphi_p$ is the pump phase.

By applying a modulated voltage to the embedded varactors in each SRR, we change their capacitance. The $C(U)$ characteristics of all real varactors are nonlinear, and we consider the exact dependence in the numerical simulations in this study. For analytical considerations we assume that the capacitance also changes according to the simple cosine periodic law. If the pump amplitude is much greater than the signal amplitude ($U_p \gg U_s$), then it is possible to neglect the capacitance change under the action of the incident wave signal, and we end up with the expression:

$$C(t) = C_0[1 - \beta \cos(\omega_p t + \varphi_p)], \quad (7)$$

where $C_0$ is the average value of the capacitance, $\beta = \frac{\Delta C}{C_0}$ is the parametric modulation coefficient, $\Delta C$ is the amplitude of the capacitance parametric modulation.

The charge on the varactor, as well as the current and voltage, are related by the following expressions:

$$q(t) = C(t)u_C(t), \quad (8)$$

$$i(t) = \frac{dq}{dt} = C(t)\frac{du_C}{dt} + u_C(t)\frac{dC}{dt}. \quad (9)$$

The linear parametric capacitance $C(t)$ includes the pump source action. Therefore, $u_C(t) = u(t)$ and the total current through capacitance $C(t)$ can be expressed as:

$$i(t) = -\omega_s C_0 U_s \sin(\omega_s t) +$$
$$+ \frac{1}{2}(\omega_p + \omega_s)\Delta C U_s \sin[(\omega_p + \omega_s)t + \varphi_p] + \quad (10)$$

$$+\frac{1}{2}(\omega_p - \omega_s)\Delta C U_s \sin[(\omega_p - \omega_s)t + \varphi_p].$$

If $\omega_p = 2\omega_s$, then only the first and third terms of Eq. 10 contribute to the total current oscillating with frequency $\omega_s$, because the second component oscillates outside the band of interest (at $3\omega_s$). The first term is the response of the constant capacitor $C_0$. As shown below, the third term can be represented by the effective conductance $G_{eq}$ connected in parallel to $C_0$. The third current component in Eq. 10 at $\omega_s$ is:

$$i_{\omega_s}(t) = \frac{1}{2}\omega_s \Delta C U_s \sin(\omega_s t + \varphi_p) = \frac{1}{2}\omega_s \Delta C U_s \cos\left(\omega_s t + \varphi_p - \frac{\pi}{2}\right). \quad (11)$$

For some phases of the pump source and incident wave, the parametric capacitance can generate or amplify currents at $\omega_s$, which can be radiated by the SRR or feed the currents excited by the incident wave in the case of parametric amplification. The power delivered to the incident wave can be calculated as follows:

$$P_{\omega_s} = \frac{U_s I_{\omega_s}}{2}\cos\left(\varphi_p - \frac{\pi}{2}\right) = \frac{1}{2}\omega_s \Delta C \frac{U_s^2}{2}\sin\varphi_p = \frac{G_{eq} U_s^2}{2}, \quad (12)$$

where $G_{eq} = \frac{\beta \omega_s C_0 \sin\varphi_p}{2}$ is the equivalent active conductance and $I_{\omega_s} = \frac{1}{2}\omega_s \Delta C U_s$ is the current amplitude. For $\varphi_p = -\frac{\pi}{2}$ the equivalent active conductance is negative: $G_{eq} = -\frac{\beta \omega_s C_0}{2}$. Thus, the total conductance of the circuit can be reduced or even become negative. Therefore, the variable capacitance added to the SRR amplifies the currents induced by the incident wave.

To simplify the analysis, we replace the lumped element circuit with an equivalent dual circuit applying the principle of duality [40,41]. This circuit includes a current source loaded by the SRR and varactor impedances, as illustrated in Figure 2. Note that the dual equivalent inductance has the value $C_0$ and the dual equivalent capacitance - $L_{SRR}$, respectively. The overall capacitance $C_0 = C_{SRR} + C_{0\,var} + C_f$ includes the capacitance of the split-ring resonator $C_{SRR}$, the filtering capacitance $C_F$ and the average capacitance of the varactor diode $C_{0\,var}$ corresponding to the value when a constant bias voltage $U_0$ is applied.

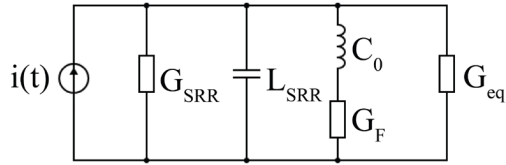

Fig. 2. An equivalent dual circuit containing a current source loaded by SRR and varactor impedances. Note that the values of inductance and capacitance are defined by the values of capacitance and inductance, respectively, in the original circuit.

If there is no parametric modulation, the output voltage amplitude of the source is $U_s = \frac{I}{G_{SRR}+G_{add}}$, where $G_{add} = \frac{C_0^2 G_F \omega_s^2}{G_F^2 + C_0^2 \omega_s^2}$, $G_F = \frac{1}{R_F}$, $G_{SRR} = \frac{1}{R_{SRR}}$ and $I$ is the current source amplitude ($G_{add} \approx G_F$ for our specific circuit parameters).

The active re-radiated power by the SRR can be determined as follows:

$$P_s = \frac{1}{2}Re\{U_S I^*\} = \frac{1}{2}Re\{U_S(U_S^* G_{SRR})\} = \frac{1}{2}U_S^2 G_{SRR} = \frac{1}{2}\frac{I^2 G_{SRR}}{(G_{SRR}+G_{add})^2}, \quad (13)$$

where $U_S$ and $I$ denote voltage and current in phasor form, * denotes complex conjugation.

In the case of parametric modulation and $G_{eq} < 0$, the procedure is similar to the one described above, but the source output voltage becomes $U_s = \frac{I}{G_{SRR}+G_{add}+G_{eq}}$. Therefore, the re-radiated power in the parametrically modulated system can be found as:

$$P_s' = \frac{1}{2}\frac{I^2 G_{SRR}}{(G_{SRR}+G_{add}+G_{eq})^2}. \tag{14}$$

Then the power gain (compared to the SRR without any pump) is:

$$K_P = \frac{P_s'}{P_s} = \frac{1}{\left(1+\frac{G_{eq}}{G_{SRR}+G_{add}}\right)^2}. \tag{15}$$

When the absolute value of the equivalent active conductance $|G_{eq}|$ in the system exceeds the total conductance of the circuit, the resonator becomes unstable, and it starts oscillating even when there is no incident wave at the signal frequency. The self-excitation threshold of the parametric amplifier or parametric generation condition can be determined from the following condition:

$$|G_{eq}| > G_{SRR} + G_{add}. \tag{16}$$

This implies that parametric generation appears in the SRR with the equivalent active conductance exceeding the total conductance of the circuit. The critical value of the parametric modulation coefficient is

$$\beta_{critical} = \frac{2\,G_o}{\omega_s C_0} = \frac{2}{Q}, \tag{17}$$

where Q is the quality factor of the SRR including all losses $G_o = G_{SRR} + G_{add}$.

The total conductance of the circuit is determined by the bandwidth of the resonance and the overall average capacitance:

$$G_o = \Delta\omega C_0. \tag{18}$$

The critical capacitance modulation value at which the instability occurs is then estimated as

$$\Delta C_{critical} = \frac{2G_o}{\omega_s} = \frac{2\Delta\omega C_0}{\omega_s}. \tag{19}$$

## 3. Full-wave numerical simulations

Next, we proceed to full-wave numerical simulations of the proposed metasurface using CST Studio. We perform numerical simulations of a periodic structure using finite-element method and output the results of 3D modeling into Schematic simulator to model the nonlinear lumped element circuit that includes the varactor diode and the biasing network. We employ the Spectral Lines task and the varactor SPICE model [42] within Schematic in CST Studio to conduct simulations of the nonlinear circuit in frequency domain. This task specifically facilitates Harmonic Balance (HB) analysis, enabling us to examine the pump and signal mixing processes within the designed circuit. This method encompasses mixing frequencies up to a specified limit of harmonics. To ensure accurate representation of the mixing effect, we utilize 20 harmonics for the signal source and 10 harmonics for the pump source, optimizing our approach for the most precise results.

Figure 3 shows the calculated transmission and reflection coefficients for the incident linearly polarized electromagnetic wave for different values of the DC voltage applied to the varactor diodes ($U_0$). We set the operating bias voltage $U_0 = 15$ V and modulate the varactor capacitance using the AC pump source as discussed earlier.

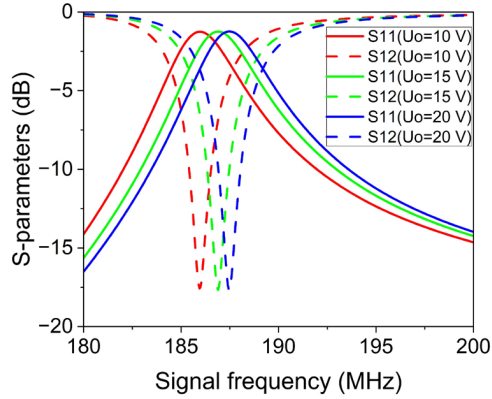

Fig. 3. Transmission (S12) and reflection (S11) of the electromagnetic wave incident on the metasurface for one of the polarizations and different values of the applied DC voltage.

The critical capacitance modulation value $\Delta C_{critical}$ can be found as described above and is approximately 0.4 pF. This corresponds to the pump amplitude of $U_{p\ critical} \approx 11$ V when the DC bias voltage is 15 V for the considered varactor diode (SMV1405). Above this pump amplitude, the circuit becomes unstable and self-oscillating. As noted above, the capacitance-voltage characteristic of the varactor is not linear. Therefore, when we drive the varactor diode with a cosine bias voltage, $C(t)$ does not follow a cosine function. This explains the small difference in the threshold values predicted by the analytical model and full wave numerical simulations. The self-excitation threshold was validated by modeling the proposed metasurface using the Spectral Lines task in CST Studio. We interpret self-oscillations as indicating the instability of the circuit for the purpose of parametric amplification. We observe that the power of self-oscillations grows exponentially with pump amplitude over $E_{p\ critical}$. This appears in the presence of the pump in the circuit and in the absence of incident electromagnetic waves. The self-excitation threshold of the parametric amplifier or the parametric generation condition corresponds to a change in the resonance frequency of the circuit that is equal to its bandwidth [40,43].

Parametric amplifiers operate substantially differently in the degenerate regime, when $\omega_p = 2\omega_s$, and in the non-degenerate regime, when $\omega_p \neq 2\omega_s$. In the degenerate regime, parametric amplifiers require strict phase relationships between the signal and pump. The maximum gain in the degenerate regime occurs in the pump phase when the equivalent conductance (see Eq. 12) of the varactor reaches maximum negative value, and the pump energy is transferred to the induced currents at $\omega_s$. When the inserted resistance (equivalent conductance) of the varactor becomes positive, parametric modulation introduces additional losses in the degenerate case. In contrast, the gain in the non-degenerate regime does not depend on the pump phase, but the phases of the reflected and transmitted waves depend on the phase of the pump. This is similar to the results reported in Ref. 24, where it was demonstrated that the phases of the pump waves control the phase of the transmitted and reflected signals.

In addition to the phase dependence, the gain changes with the pump amplitude. When approaching the instability threshold, the gain increases with an increase in the pump amplitude, as shown in Figure 4. The maximum gain reaches approximately 16 dB in the degenerate regime (under conditions of phase locking) and 10 dB in the non-degenerate regime when the pump with the amplitude $U_p = 11$ V at the frequency of 372 MHz is applied to split-ring resonators and the small-amplitude wave is incident on the metasurface at 186 and 186.1 MHz, respectively. The results are shown up to $U_p = 11$ V because this value corresponds to the instability threshold of the system described previously. Figure 4(a) also shows the theoretically predicted overall gain calculated using Eq. 15. As the exact ratio of the incident power and the induced electromotive force in the equivalent circuit cannot be found analytically with sufficient accuracy, we have normalized the obtained values, so that the gain without the pump ($U_p = 0$ V) matches the value obtained numerically.

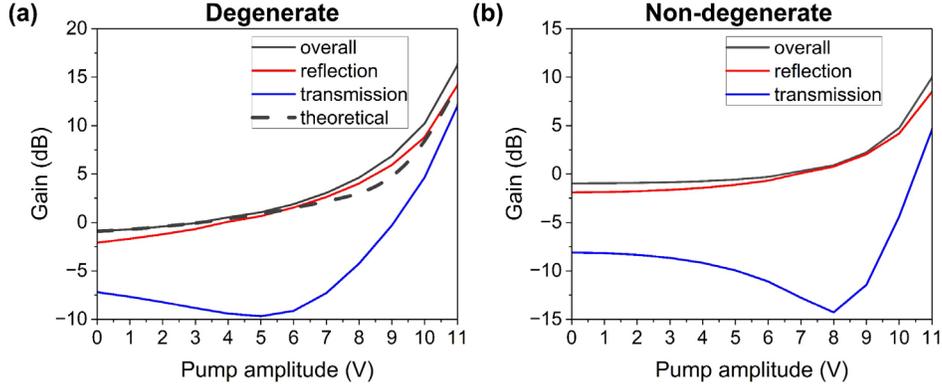

Fig. 4. The dependence of the gain on the pump amplitude in the degenerate $\omega_s = 186$ MHz, $\omega_p = 372$ MHz (a) and non-degenerate $\omega_s = 186.1$ MHz, $\omega_p = 372$ MHz (b) cases. The dashed line shows the theoretically predicted overall gain from Eq. 15.

In both the degenerate and non-degenerate cases, the reflected signal and the total scattered energy continuously increase with the pump amplitude. The signal transmitted through the metasurface first decreases as the pump amplitude increases to a certain threshold value, after which it also begins to increase. This occurs because the selected operating frequency is below the transmission resonance. Due to the nonlinearity of the capacitance-voltage characteristic, the average capacitance over a wave period increases with a rise in pump amplitude, similar to the results reported in Ref. 44. The rise in capacitance results in a decrease in the resonant frequency of the circuit. Consequently, for frequencies below the resonance frequency when $U_0 = 15$ V, this initiates a reduction of the transmission coefficient. However, the parametric gain experiences exponential growth with increasing pump amplitude, eventually reaching a level sufficient to offset the decline in the transmission coefficient. This explains the initial drop in transmission, where the coefficient decreases while the parametric effect remains weak, unable to fully compensate for the decline. As the pump amplitude continues to rise, the parametric gain eventually surpasses the decrease in transmission, leading to an overall increase in gain.

The dependence of the metasurface gain on the incident wave frequency in the degenerate case is shown in Figure 5(a). It is calculated in the regime of phase-locking of the pump source with the small-amplitude incident wave ($U_s = 0.01$ V) and the pump amplitude $U_p = 10$ V, which is below the instability threshold. We maintain the degenerate condition $\omega_p = 2\omega_s$ by changing both the pump and signal frequencies in this regime. This plot illustrates that maximum amplification occurs at frequencies slightly lower than the resonance frequency when the DC bias voltage is $U_0 = 15$ V (indicated by dotted vertical lines), approximately around 186 MHz.

Figure 5(b) shows the dependence of the gain on the signal frequency in the non-degenerate case for the pump amplitude $U_p = 10$ V. The gain does not depend on the phase; therefore, there is no need for phase-locking. The magnitude of the overall gain is slightly lower than the maximum possible gain in the degenerate case shown in the middle ($\omega_s = 186$ MHz, $\omega_p = 372$ MHz). The bandwidth of degenerate amplification is exceedingly narrow; hence, in practice, utilizing this regime may pose challenges. With any frequency mismatch, some of the energy is converted to the idler frequency, thereby reducing the gain. We observe that the overall achievable gain is higher in the degenerate case. Most remarkably, the amplification mostly occurs for the reflected wave in the non-degenerate case, whereas the transmitted wave is further suppressed from its already low value near the resonance.

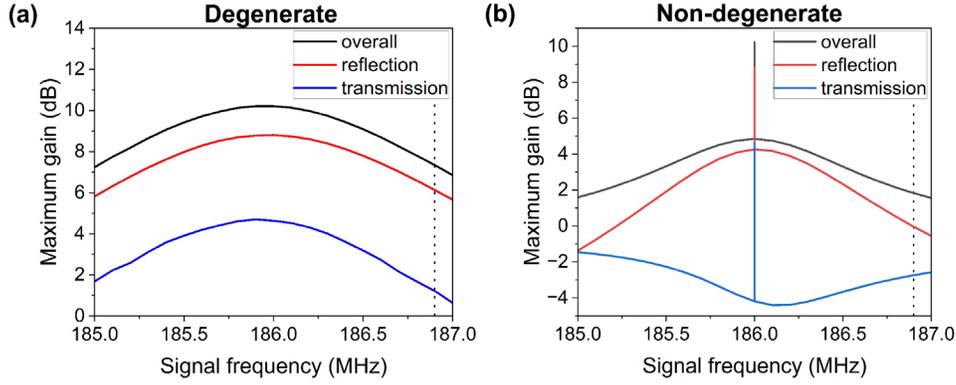

Fig. 5. The dependence of the gain (overall, for reflection and transmission separately) on the frequency of the electromagnetic wave incident on the metasurface in the degenerate case (a), where we maintain $\omega_p = 2\omega_s$, and the non-degenerate case (b), where $\omega_p = 372$ MHz. The dotted vertical line shows the resonance position when $U_0 = 15$ V.

A comparison of the dependence of the gain on the signal amplitude in the degenerate and non-degenerate regimes for the pump amplitude $U_p = 8$ V is shown in Figure 6. Amplification shows no signs of saturation, and the gain is restricted by the breakdown voltage of the varactor diode in both cases. The signal amplitude becomes sufficiently high to significantly modulate the pump, so it reaches the breakdown value when $U_s \approx 1.5$ V.

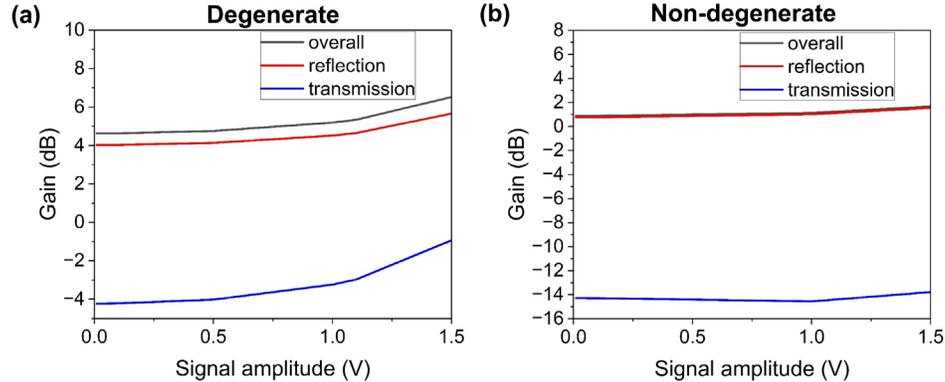

Fig. 6. The dependence of the gain on the signal amplitude in the degenerate (a) and non-degenerate (b) cases.

## 4. Discussion

The obtained simulation results are consistent with the theory of parametric generation and amplification and the developed analytical model. We demonstrate a significant gain of more than 10 dB in both the degenerate and non-degenerate regimes for a small incident radiation power (tens of nanowatts). These results were confirmed for the incident radiation power up to approximately 0.75 mW. The theoretically calculated self-excitation threshold matches the results of numerical simulations of the parametric metasurface. The proposed amplification method can be used in the radio, microwave, and subterahertz ranges and is limited by the properties of modern varactor diodes (up to about 500 GHz).

The disadvantage of degenerate parametric amplification is the need for phase locking of the pump source with the incident wave. In practice, this can only be realized if there is information about the phase of the incident radiation and if the frequency of the pump source is fine-tuned to the condition $\omega_p = 2\omega_s$. Therefore, the proposed parametric metasurface is more practically useful in the non-degenerate regime.

Further improvements to the proposed metasurface include better filtering circuits or the use of non-radiating modes of resonators to avoid radiation at the pump frequency. It is also possible to use the pump without electrically applying parametric modulation but inducing currents in the resonators at $\omega_p \approx 2\omega_s$ by the incident electromagnetic wave. This requires the development of two-mode resonators to effectively couple the incident radiation.

We anticipate that the proposed structure can be used not only for amplifying electromagnetic waves but also for providing spatial control of the scattered waves by employing spatially inhomogeneous amplification. This integration of multiple functions can be useful in various applications, including multifunctional and programmable radomes or ground and orbit-based repeaters that can amplify incident signals.

Given that the scatterers in this study primarily reflect incident radiation at the resonance frequency, it is inherent that the amplified scenario is dominated by reflected waves. We anticipate that employing a structure primarily designed for resonant forward radiation scattering would lead to a prevalence of transmitted waves.

These amplification techniques can be applied to quantum communications, such as the development of low-noise repeaters required for quantum channels over long distances. Given that parametric amplification finds extensive use in readouts for quantum computing, the proposed parametric metasurfaces hold the potential to enhance existing technologies in this area.

Future directions include the development of a metasurface based on two-circuit parametric amplifiers that can eliminate the shortcomings of one-circuit non-degenerate parametric amplifiers, such as beating. Other tuning methods that can modulate the parameters of metasurfaces at higher rates can extend these amplification techniques to the terahertz and infrared ranges.

## 5. Conclusion

In summary, our study reveals the potential of amplifying electromagnetic waves through the application of a variable capacitor-loaded metasurface. Our investigation successfully demonstrates the capability to achieve significant gain exceeding 10 dB. The robust validation of the developed theoretical model, conducted through comprehensive full-wave numerical simulations, underscores the reliability and applicability of the proposed amplification method. This research provides valuable insights into the dynamic field of time-varying metasurfaces, paving the way for advancements in the manipulation of electromagnetic waves.

**Funding.** This research was supported by the Australian Research Council Centre of Excellence for Transformative Meta-Optical Systems (Project ID CE200100010).

**Acknowledgments.** The authors thank A.B. Kozyrev for useful discussions.

**Disclosures.** The authors declare no conflicts of interest.

**Data availability.** Data underlying the results presented in this paper are not publicly available at this time but may be obtained from the authors upon reasonable request.